\documentclass[conference]{IEEEtran}

\usepackage{amsmath,amssymb,amsfonts}
\usepackage{algorithmic}
\usepackage{graphicx}
\usepackage{textcomp}
\usepackage{xcolor}
\usepackage[font=footnotesize]{caption}
\usepackage{subcaption}
\usepackage{./resources/tikz/tikzit}

\tikzstyle{Rectangle Medium}=[fill={rgb,255: red,42; green,71; blue,101}, draw=black, shape=rectangle, minimum width=2cm, text=white, minimum height=0.75cm, align=center]
\tikzstyle{Rectangle Large}=[fill={rgb,255: red,42; green,71; blue,101}, draw=black, shape=rectangle, text=white, minimum width=4cm, minimum height=2cm, align=center]
\tikzstyle{Red Text}=[fill=none, draw=none, shape=rectangle, tikzit draw={rgb,255: red,255; green,131; blue,131}, tikzit fill={rgb,255: red,255; green,131; blue,131}, text={{rgb,255: red,255; green,131; blue,131}}, align=center]
\tikzstyle{Small Circle}=[fill={rgb,255: red,42; green,71; blue,101}, draw=black, shape=circle, text=white]
\tikzstyle{Text}=[fill=none, draw=none, shape=rectangle, tikzit draw=black, align=center]
\tikzstyle{White Text}=[fill=none, draw=none, shape=rectangle, tikzit draw=black, align=center, text=white]
\tikzstyle{Rectangle Small}=[fill={rgb,255: red,42; green,71; blue,101}, draw=black, shape=rectangle, text=white, align=center, minimum height=0.5cm, minimum width=1.5cm]
\tikzstyle{Rectangle MedLarge}=[fill={rgb,255: red,42; green,71; blue,101}, draw=black, shape=rectangle, text=white, align=center, minimum height=1.33cm, minimum width=2.5cm]
\tikzstyle{Small Outline Circle}=[fill=white, draw=black, shape=circle, inner sep=0pt]

\tikzstyle{Arrow}=[draw=black, ->, line width=.025cm]
\tikzstyle{Edge}=[-, draw=black, line width=0.025cm, fill=none]
\tikzstyle{Red Dashed}=[-, draw={rgb,255: red,255; green,131; blue,131}, dashed, line width=0.025cm]
\tikzstyle{Dashed}=[-, draw={rgb,255: red,136; green,136; blue,136}, dashed, line width=0.025cm]
\tikzstyle{Freebox}=[-, draw=black, fill={rgb,255: red,42; green,71; blue,101}]
\tikzstyle{Freebox Green}=[-, draw=black, fill={rgb,255: red,115; green,167; blue,144}]
\tikzstyle{Freebox Yellow}=[-, draw=black, fill={rgb,255: red,215; green,177; blue,124}]
\tikzstyle{Freebox Red}=[-, draw=black, fill={rgb,255: red,255; green,131; blue,131}]
\tikzstyle{Thick Dashed}=[-, line width=0.025cm, draw=black, dashed]
\tikzstyle{Black Lines}=[-, fill=none, pattern=north east lines, pattern color={{rgb,255: red,42; green,71; blue,101}}, draw=black, tikzit fill={rgb,255: red,191; green,191; blue,191}, tikzit draw={rgb,255: red,241; green,239; blue,228}, line width=0.025cm]
\tikzstyle{Blue Background}=[-, fill={{rgb,255: red,42; green,71; blue,101}}, fill opacity=0.3, draw=none]
\tikzstyle{Green Background}=[-, fill={{rgb,255: red,115; green,167; blue,144}}, fill opacity=0.45, draw=none]
\tikzstyle{Black Lines Dashed}=[-, fill=none, pattern=north east lines, pattern color={{rgb,255: red,42; green,71; blue,101}}, draw=black, tikzit fill={rgb,255: red,191; green,191; blue,191}, tikzit draw={rgb,255: red,241; green,239; blue,228}, line width=0.025cm, dashed]
\tikzstyle{Blue Dots}=[-, fill=none, pattern=horizontal lines, pattern color={{rgb,255: red,215; green,177; blue,124}}, tikzit fill={rgb,255: red,42; green,71; blue,101}, line width=0.025cm]
\tikzstyle{Red Bricks}=[-, fill=none, pattern=crosshatch, line width=0.025cm, tikzit fill={rgb,255: red,255; green,131; blue,131}, pattern color={{rgb,255: red,115; green,167; blue,144}}]
\tikzstyle{Black Cross}=[-, fill=none, pattern=crosshatch dots, line width=0.025cm, tikzit fill={rgb,255: red,255; green,131; blue,131}, pattern color={{rgb,255: red,42; green,71; blue,101}}]
\tikzstyle{Red Bricks Dashed}=[-, fill=none, pattern=crosshatch, line width=0.025cm, tikzit fill={rgb,255: red,255; green,131; blue,131}, pattern color={{rgb,255: red,115; green,167; blue,144}}, dashed]
\tikzstyle{Dashed Arrow}=[->, line width=0.025cm, draw=black, dashed]
\tikzstyle{Thin Dashed Arrow}=[->, line width=0.025cm, dashed, draw={rgb,255: red,136; green,136; blue,136}]
\tikzstyle{Blue EC F}=[-, draw={rgb,255: red,42; green,71; blue,101}, fill={rgb,255: red,42; green,71; blue,101}, fill opacity=0.33, line width=0.25mm]
\tikzstyle{Green EC F}=[-, draw={rgb,255: red,115; green,167; blue,144}, fill={rgb,255: red,115; green,167; blue,144}, fill opacity=0.33, line width=0.25mm]
\tikzstyle{Yellow EC F}=[-, draw={rgb,255: red,215; green,177; blue,124}, fill={rgb,255: red,215; green,177; blue,124}, fill opacity=0.33, line width=0.25mm]
\tikzstyle{Pink EC F}=[-, draw={rgb,255: red,234; green,186; blue,185}, fill={rgb,255: red,234; green,186; blue,185}, fill opacity=0.33, line width=0.25mm]

\usepackage{svg}
\usepackage[style=ieee]{biblatex}
\bibliography{bib/refs.bib}

\usepackage{flushend}
\usepackage{booktabs}
\usepackage{enumitem}

\DeclareUnicodeCharacter{2212}{-}

\def\BibTeX{{\rm B\kern-.05em{\sc i\kern-.025em b}\kern-.08em
    T\kern-.1667em\lower.7ex\hbox{E}\kern-.125emX}}

\newcommand{\Description}[1]{}

\newcommand\copyrighttext{%
  This work has been submitted for possible publication. Copyright may be transferred without notice, after which this version may no longer be accessible.
}

\newcommand\copyrightnotice{%
\begin{tikzpicture}[remember picture,overlay]
\node[anchor=south,yshift=20pt] at (current page.south) {\fbox{\parbox{\dimexpr\textwidth-\fboxsep-\fboxrule\relax}{\copyrighttext}}};
\end{tikzpicture}%
}

\begin{document}

\title{Beyond the Limits: Flexible and Congestion-Aware Cluster Scheduling for the Cloud}

\makeatletter
\newcommand{\linebreakand}{%
  \end{@IEEEauthorhalign}
  \hfill\mbox{}\par
  \mbox{}\hfill\begin{@IEEEauthorhalign}
}
\makeatother

\author{

\IEEEauthorblockN{Oliver Larsson}
\IEEEauthorblockA{Dept. of Computing Science\\
Ume\aa{} University, Sweden\\
olars@cs.umu.se}
\and
\IEEEauthorblockN{Thijs Metsch}
\IEEEauthorblockA{Unaffiliated\\
Germany\\
tmetsch@engjoy.eu}
\and
\IEEEauthorblockN{Cristian Klein}
\IEEEauthorblockA{Dept. of Computing Science\\
Ume\aa{} University, Sweden\\
cklein@cs.umu.se}
\and
\IEEEauthorblockN{Erik Elmroth}
\IEEEauthorblockA{Dept. of Computing Science\\
Ume\aa{} University, Sweden\\
elmroth@cs.umu.se}

}

\maketitle

\copyrightnotice

\begin{abstract}
    Workload scheduling in cloud environments often relies on simplistic assumptions 
about application resource needs and hardware utilization. 
Overlooking application-level performance objectives and hardware resource contention that leads
to inefficient resource usage and degraded performance. 
This paper addresses two key limitations of current approaches. First, 
unnecessarily strict enforcement of service level objectives (SLOs) often leads 
to resource underutilization and poor energy efficiency. Second, lack of congestion awareness in shared 
resources such as last-level cache (LLC) and memory bandwidth.
In this paper, we propose two complementary strategies to address these limitations: 
(i) integrating soft SLO limits that allow controlled overcommitment and 
tolerate minor, transient violations to improve cluster efficiency, and (ii) 
introducing resource-aware scheduling and rescheduling based on real-time 
congestion insights for shared resources such as last-level cache (LLC) 
and memory bandwidth. 
Our results show that soft SLO limits reduce corrective rescheduling actions by 49\% compared to hard-limit 
approaches while maintaining acceptable performance guarantees. Additionally, 
resource-aware scheduling decreases node-level congestion by 8\% and further mitigates SLO 
violations, demonstrating the effectiveness of incorporating 
application-level flexibility and hardware-level insights into scheduling 
and rescheduling decisions.

\end{abstract}


\newcommand\comment[1]{}
\renewcommand\comment[1]{[\textcolor{red}{#1}]}

\section{Introduction}

Cluster computing systems are essential to modern data processing, 
enabling scalable and resilient execution of workloads across datacenters and beyond. 
Effective resource management and placement
of workloads is critical to maintaining performance, efficiency, and 
resiliency~\cite{tirmaziBorgNextGeneration2020,zhangZeusImprovingResource2021a}. 
Kubernetes,\footnote{\url{https://kubernetes.io}} the de facto standard for 
container orchestration, employs a heuristic scheduling approach: workloads are 
placed on nodes based on static, user-defined resource requests and 
current node allocations~\cite{carrionKubernetesSchedulingTaxonomy2022,rejibaCustomSchedulingKubernetes2022}. 
While simple and efficient, this greedy strategy ignores application-level 
performance metrics and resource characteristics beyond CPU and memory.

In practice, application owners care more about application-level performance and service level objectives (SLOs)
than about raw resource quantities, a principle captured by intent-driven 
orchestration~\cite{sharma_sla_2023}. Furthermore, congestion in shared resources 
such as the last-level cache (LLC) and memory bandwidth can significantly 
degrade performance~\cite{sohalCloserLookIntel2022,larssonHardwareLevelQoSEnforcement2025}, 
yet current schedulers provide no mechanism for expressing such resource requirements. 
Consequently, current heuristic scheduling approaches often result in suboptimal workload placements, 
harming application performance and energy efficiency due to suboptimal hardware utilization~\cite{larssonImpactDirectedPod2023}. 

This paper addresses two key limitations we have identified in scheduling approaches for cloud native environments:
\begin{enumerate}
\item Current SLO-based scheduling treats objectives as hard limits, often leaving cluster 
    resources underutilized to accommodate transient load spikes without violating SLOs.
\item Existing approaches fail to leverage hardware-level insights, such as LLC and memory bandwidth contention, 
    to inform scheduling and rescheduling decisions. 
\end{enumerate}

\subsubsection{Soft SLO Limits}

Several studies have enhanced the scheduling process by incorporating application-level
metrics and actual performance numbers into the decision-making 
process~\cite{zhangZeusImprovingResource2021a,marcheseEnhancingKubernetesPlatform2025,fuQoSAwareResourceEfficient2021,carvalhoQoEAwareContainerScheduler2021}.
Many such approaches also enable rescheduling of workloads based on dynamic performance metrics and
resource usage patterns~\cite{fuQoSAwareResourceEfficient2021,larssonImpactDirectedPod2023}.
Even so, most approaches fail to consider the service level
indicators (SLIs) and SLOs of the workloads being scheduled.
Those who do (e.g. \citeauthor*{da_silva_qos-driven_2020}~\cite{da_silva_qos-driven_2020}) 
treat SLOs as hard limits, expecting an immediate penalty upon breach. While this may be 
suitable in certain scenarios, it carries the risk of enforcing cluster underutilization
that leads to poor energy efficiency
in the real world due to the inherent variability in workloads and hardware performance, especially under
peak loads and transient spikes. 
Therefore, instead of taking immediate corrective action, we here investigate the benefits 
and consequences of using ``soft'' SLO limits with defined margins that allow 
controlled overcommitment and brief, minor violations, reducing overhead 
while preserving acceptable performance guarantees.

\subsubsection{Scheduling with Resource Insight}

Previous studies have demonstrated the negative impact of congestion over
shared system resources such as LLC and memory bandwidth on
application performance~\cite{sohalCloserLookIntel2022,larssonHardwareLevelQoSEnforcement2025}. 
To the best of our knowledge, no existing scheduling approaches leverage insights 
into current hardware resource congestion to inform scheduling decisions in cloud native environments.
Therefore, we propose a scheduling and rescheduling approach that
leverages individual resource pressure insights, enabling informed decisions
about workload placement and migration based on current contention levels in 
individual resources. To achieve this, we build upon the concept of workload buoyancy,
a framework for capturing application resource sensitivity insights at runtime~\cite{larsson_workload_2026}.

\subsection{Summary of Contributions}

In this paper, we address two key gaps in existing cluster scheduling approaches for cloud
environments. We propose complementary scheduling and rescheduling strategies that leverage
soft SLO limits and workload buoyancy to improve application-aware workload scheduling 
and to reduce resource congestion and its negative impact on application performance and energy efficiency.
We make the following contributions:

\begin{itemize}
    \item We integrate soft SLO limits into the scheduling and rescheduling
        process in orchestrated container environments and investigate its 
        effects on cluster efficiency and application performance.
    \item We introduce a novel scheduling and rescheduling approach that leverages shared
        resource congestion insights through workload buoyancy, enabling informed decisions about workload
        placement and migration based on current contention levels in individual hardware resources.
    \item Our approaches are implemented as extensions to the Kubernetes scheduler and
        descheduler, demonstrating their practical applicability in cloud native environments.
\end{itemize}

Through a comprehensive experimental evaluation using representative workloads, we demonstrate
the effectiveness of our proposed approaches. 
Using our soft SLO limit approach, we reduce the average SLO violation size 
compared to a baseline Kubernetes configuration, while using 49\% fewer corrective
rescheduling actions compared to a hard SLO limit approach. Additionally, by incorporating resource congestion
awareness into scheduling decisions, we reduce overall node resource congestion by 8\% while
reducing SLO violations.

\section{Background}

Scheduling workloads is a critical aspect of modern computing systems. 
Efficient placement directly impacts resource utilization, energy efficiency, 
and system performance~\cite{carrionKubernetesSchedulingTaxonomy2022,rejibaCustomSchedulingKubernetes2022}. 

\subsection{Cloud Native Scheduling and Descheduling} \label{sec:background-scheduling}

Kubernetes uses a typical two-phase scheduling process for workloads~\cite{larssonImpactDirectedPod2023}. 
After a workload is created, the scheduler decides which cluster node should host and run the workload.
These workloads, called Pods, are scheduled individually in isolation. 
The first phase, filtering, eliminates nodes that are deemed unfeasible for the Pod
based on various constraints such as resource requests, node taints, and affinity rules.
Should no nodes be feasible targets, the Pod remains unscheduled until a suitable node becomes available.

The second phase, scoring, ranks all feasible nodes using various scoring functions
to determine the most suitable node. The node with the highest score
is selected as the target for the Pod, which is then bound to that node.
A node's score is typically based on resource availability, 
load balancing, and availability considerations~\cite{carrionKubernetesSchedulingTaxonomy2022}.

Kubernetes does not support any form of rescheduling natively. However, the
\emph{descheduler}\footnote{\url{https://github.com/kubernetes-sigs/descheduler}} is an add-on component
that provides similar functionality. It periodically evaluates the cluster state
and identifies Pods that would benefit from being evicted and rescheduled.
Such Pods are deleted from their current node. This triggers the managing
replication controller to create a new instance of the Pod, which is scheduled
once again, effectively allowing the scheduler to make a new
placement decision under current cluster conditions~\cite{larssonImpactDirectedPod2023}.

Both the scheduler and descheduler are extensible through extension 
points that can modify their behavior during filtering and scoring.
This enables the implementation of custom scheduling and descheduling strategies
that can leverage additional insights and metrics beyond the default ones~\cite{larssonImpactDirectedPod2023}.

\subsection{Workload Slack and Margins} \label{sec:background-slack}

In the domain of applications, performance is typically quantified using 
SLIs such as tail latency, throughput, or availability,
depending on the nature of the application. In many cases, targets or limits 
are set in the form of SLOs that define acceptable
SLI performance thresholds. The performance \emph{slack} of a workload is 
a measure of the headroom available during operation.
It is often defined as the difference between the actual performance and the 
SLO target in some SLI~\cite{loHeraclesImprovingResource2015}.
Therefore, a workload with high slack can tolerate more resource contention
without violating its SLOs, while a workload with low or negative slack 
is more sensitive to additional load and resource contention as it is close
to breaching, or have already breached, its SLO.

Generally, workloads will operate with a certain degree of slack such 
that they can handle transient spikes in demand or resource contention without
breaching their SLOs. This means that there are generally resources 
and performance overhead available, left underutilized during normal operation
to ensure reliable performance under spiking conditions. Cluster operators 
aim to minimize this waste by operating workloads closer to their limits without 
jeopardizing their SLO compliance~\cite{chenPARTIESQoSAwareResource2019}.

\subsection{Resource Awareness and Workload Buoyancy} \label{sec:background-buoyancy}

In the heterogeneous and dynamic environments of modern cloud computing,
workloads exhibit varying degrees of sensitivity to different resources.
Some will be limited by CPU, others by memory bandwidth, cache size, network I/O,
or other factors~\cite{patelCLITEEfficientQoSAware2020}. For example, a plethora of research
has shown that efficient allocation of last-level cache (LLC) can have a significant
impact on application performance~\cite{chenPARTIESQoSAwareResource2019, 
zhangZeusImprovingResource2021a, larssonHardwareLevelQoSEnforcement2025}.

Only a very limited set of studies have explored the idea of scheduling based
on such a broader set of resource considerations~\cite{carrionKubernetesSchedulingTaxonomy2022}.
This can be partly attributed to the inherent complexity of accurately modeling
and predicting workload behavior under varying resource conditions, 
including challenges associated with profiling and monitoring resource usage.  

\subsubsection{Workload Buoyancy}

The concept of \emph{workload buoyancy} was recently proposed by \citeauthor*{larsson_workload_2026} as
a measure of how well a workload can maintain its performance in 
the face of resource contention and load variability that attempts to capture
the resource characteristics of workloads and nodes in a holistic manner~\cite{larsson_workload_2026}.
A workload's buoyancy score is computed from \emph{resource scores}, 
a measure of its sensitivity to various resources,
and its performance slack. It is a composite metric that is computed online
without requiring any prior profiling of the workload or its resource usage patterns.
It is able to provide insight into how well
the resource utilization of a node can be increased without negatively impacting
its workloads. More specifically, a buoyancy score is a value in the range $(-\infty, 1]$ where
a positive value indicates that the workload has some degree of slack and 
can tolerate some additional resource contention, while a negative value indicates that
the workload is already breaching its SLOs and is highly sensitive to additional
contention. A buoyancy score close to 0 indicates that the workload has
little slack and is sensitive to contention, while a score close to 1 indicates
that the workload has high slack and is less sensitive to contention.

The buoyancy score of a workload is a composite metric that integrates
resource scores across multiple resource dimensions (e.g. CPU, LLC, memory bandwidth).
Each resource score is a value in the range $[0, 1]$, and reflects the 
workload's sensitivity to contention for that specific resource. A resource score
close to 0 indicates low sensitivity to contention for that resource,
while a score close to 1 indicates high sensitivity and a likely bottleneck.
Such scores allow for a more nuanced understanding of workload behavior,
and comparison across different workloads that exhibit different resource
usage behavior.

\section{Problem Statement} \label{sec:problem-statement}

Current cluster workload scheduling approaches primarily focus on major system
resources such as CPU and memory when making scheduling decisions. Even when 
application-level metrics such as SLIs are considered, they are typically 
incorporated in one of two ways: They are either
treated as optimization objectives to be maximized or minimized, or as hard 
constraints that must be satisfied. 
In practice, many services have performance targets that could be treated as soft 
limits, allowing for controlled tolerance of transient breaches. However, 
existing schedulers lack mechanisms to exploit this flexibility. This raises two key questions:
\begin{enumerate}
    \item What are the effects of treating SLOs as soft limits?
    \item Can such flexibility be integrated into scheduling decisions to reduce unnecessary rescheduling actions?
\end{enumerate}

Additionally, current scheduling strategies overlook the complex interactions 
between workloads and shared resources beyond CPU and memory, such as LLC and memory
bandwidth. These resources significantly influence performance in multi-tenant
environments where workloads compete for shared capacity. While workload buoyancy
has been shown to capture these interactions in node-level resource allocation 
contexts~\cite{larsson_workload_2026}, its potential for cluster-level scheduling remains 
unexplored. This leads to yet another question:

\begin{enumerate}
    \setcounter{enumi}{2}
    \item Can workload buoyancy and its resource score components be effectively leveraged to improve scheduling decisions in cloud native environments?
\end{enumerate}

\subsubsection{Considerations}

Scheduling and rescheduling of workloads is one way to address SLO breaches caused by resource 
contention, but it is not the only solution. In scenarios involving increased load, 
application-level mechanisms such as autoscaling (e.g. horizontal pod autoscaling)
may also be effective. However, such approaches solve the problem by increasing 
the resources available to the workload, which may not be possible or desirable
from the perspective of energy and cost. We show that issues can be mitigated
without strictly increasing workload resources available.

\section{Solution Design}

To address the outlined challenges, we propose a two-part scheduling solution. 
The first component introduces temporal stability by treating SLO violations 
as soft limits, while the second incorporates resource contention insights into 
scheduling and descheduling decisions through buoyancy awareness. 

\subsection{Soft Limits and Margins}\label{sec:soft-limits}

Under stable conditions, a workload currently in breach of its SLO will most likely
remain in breach if resource contention and workload load levels remain unchanged.
In the view of an individual worker node, rescheduling one or more workloads to another node is required to
restore SLO compliance. In real systems with varying conditions however, 
simply taking a momentary snapshot of the system state and workload performance 
may lead to excessive rescheduling actions should such breaches be transient 
in nature. If a breach is small and short-lived, it may in many cases be tolerated without 
significant impact on the user experience, while the cost of 
a corresponding rescheduling action may be non-trivial.

Therefore, we propose treating SLOs as soft limits with an associated margin.
Within the confines of this margin, SLO breaches are tolerated without triggering
rescheduling actions. Only when the breach is significant enough in relation to 
this margin, will rescheduling actions be triggered. This concept is inspired by
electrical fuses, where small overloads are tolerated for short periods of time, 
while larger overloads will cause the fuse to blow almost instantly~\cite{lee_simplified_2019}.

To evaluate this, we propose using a simple moving average to smooth out 
transient breaches, where each workload has an associated SLO slack $s_{t}$
at some time $t$. Assuming a discrete time series, we define 
the average SLO slack over a lookback window of $w$ discrete time steps up to time $t$ as
\begin{equation}
    \overline{s_{t}}(w) = \frac{1}{w} \sum_{i=0}^{w-1} s_{t-i}.
\end{equation}
We then use this average-over-time slack as basis for scheduling and rescheduling 
decisions. A larger lookback window $w$ will provide a larger margin and toleration
to transient breaches, while a smaller window will be more sensitive to small
breaches of SLO. Even with larger windows however, sustained breaches will eventually
lead to rescheduling actions as the average slack will decrease over time.
A lookback window of $w=1$ will effectively treat SLOs as hard limits. 
\figurename~\ref{fig:slack-window} illustrates the effects of different window sizes $w$.

\begin{figure}[tbp]
    \centering
    \scalebox{.5}{
        \fontsize{14}{14}\selectfont 
        \includesvg{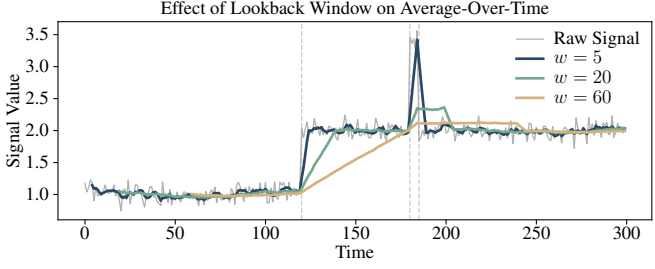}
    }
    \caption{Illustration of lookback window size on average-over-time. 
        Short windows respond rapidly to sudden changes, while longer windows 
        produce smoother but delayed responses. The raw signal value is 
        randomly generated for illustrative purposes, with major changes 
        at specified time steps.}
    \label{fig:slack-window}
\end{figure}

\subsection{Buoyancy-Aware Scheduling and Descheduling}

The resource insights buoyancy provide may be used at different phases
of the workload lifecycle. During scheduling, buoyancy metrics can inform
both the filtering and scoring phases of the scheduling process. During
filtering, node-level buoyancy scores may be used to identify unfeasible nodes that
cannot accommodate additional resource demands. Later during scoring, buoyancy 
scores may be used to rank feasible nodes based on their current resource usage
and contention levels. Nodes with larger buoyancy scores, indicating lower contention, 
may be prioritized for workload placement.

More specifically, we propose a buoyancy-aware scheduling strategy that
incorporates buoyancy metrics into the filtering and scoring phases of the scheduling process.
During filtering, we evaluate nodes based on their node-level buoyancy scores
to determine whether they can accommodate additional workloads. We do this
by setting a threshold buoyancy score, below which nodes are considered unfeasible
for new workload placements. This ensures that workloads are only scheduled
on nodes with sufficient resource availability and low contention. Here,
we experimentally select a threshold buoyancy of $0.1$, meaning that nodes
with a buoyancy score below $0.1$ are excluded from consideration during scheduling.

During scoring, feasible nodes are ranked based on their buoyancy scores. Nodes with
higher buoyancy scores, indicating lower contention,
are prioritized during scheduling. We know that the buoyancy score $B_n$ of
a node $n$ is a real number in the range $(-\infty, 1]$, where a value $< 0$ indicates
that the node is oversubscribed. In our scoring function, 
we compute the buoyancy-based score $S_n$ for each feasible node $n$ as
\begin{equation}
    S_n = 
    \begin{cases}
        \lceil B_n \times 10\rceil, &  B_n > 0 \\
        0, & B_n \leq 0.
    \end{cases}
\end{equation}
I.e., nodes with a buoyancy score $> 0$ are assigned a score in the range $[1, 10]$,
whereas nodes with a buoyancy score $\leq 0$ are assigned a score of $0$. This scoring
function ensures that nodes with higher buoyancy scores are prioritized for workload placement,
while nodes with low or negative buoyancy scores are deprioritized. With the
previous filtering step, nodes with negative buoyancy scores should
not be considered at all. However, including them in the scoring function
provides the freedom of using different threshold values during filtering, 
or to leave out the filtering completely if desired.

Furthermore, we also propose a buoyancy-aware descheduling strategy that leverages
buoyancy metrics to identify and evict workloads from oversubscribed nodes.
Initially, we use the node-level buoyancy score to identify if a node is in
a state where a descheduling action is warranted. Specifically, if a node's buoyancy
score falls below a certain threshold (e.g., $0$), it indicates that the node is oversubscribed
and would benefit from descheduling action. 
Once an oversubscribed node is identified,
its most contested resource is selected based on the node-level resource scores. 
The node's workloads are then ranked based on their individual resource scores for that specific resource, where
the workload with the largest resource score for the contested resource is evicted.

In summary, we achieve our buoyancy-aware scheduling and descheduling strategy 
by integrating buoyancy awareness in the following mechanisms: 
(1) node filtering to exclude unfeasible nodes,
(2) scoring feasible nodes, and
(3) descheduling workloads from oversubscribed nodes.

\subsection{Complementary Nature of the Two Approaches}

While both proposed mechanisms influence scheduling behavior, 
they address different aspects of cluster management and 
are intended to complement rather than replace one another. 
Soft limits is a mechanism that influences \emph{when} corrective scheduling actions should be taken, 
whereas buoyancy-aware scheduling determines \emph{how} those actions should be carried out. 
The former introduces temporal stability by preventing corrective actions in response 
to short-lived or insignificant SLO violations, while the latter improves 
the quality of corrective actions by incorporating information about 
resource contention and workload sensitivity.

Combining the two mechanisms enables the scheduler to avoid unnecessary 
corrective actions during transient performance fluctuations while still 
making informed scheduling decisions when sustained resource contention 
requires intervention. This separation of concerns allows each mechanism 
to address a distinct limitation of conventional resource-based scheduling 
without introducing conflicting objectives.

\section{Experimental Setup}

We evaluated our approach using experiments in a controlled emulated environment 
that closely replicates real-world conditions, complemented by a virtualized 
multi-node cluster simulation to assess our scheduling strategy. This section 
outlines the experimental platform, workloads, and simulation.

The experimental platform consists of a single server acting as the node under 
test within a Kubernetes cluster, running our custom scheduling and 
descheduling components. Control-plane services, monitoring, and 
load generation were isolated on separate systems to avoid interference. 
The hardware specifications of the test node are summarized in \tablename~\ref{tab:platform}.

\begin{table}[tbp]
    \centering
    \caption{Experimental platform hardware configuration.}
    \footnotesize
    \begin{tabular}{ll}
        \toprule
        \textbf{Component} & \textbf{Specification} \\ \midrule
        \textbf{CPU} & Intel Xeon Silver 4309Y\\
        \textbf{Operating System} & Ubuntu 24.04 (kern. 6.8) \\ 
        \textbf{Sockets} & 1 \\
        \textbf{Processors (Physical / Logical)} & 8 / 16 \\
        \textbf{Frequency (Base / Max Turbo)} & 2.8 GHz / 3.6 GHz \\
        \textbf{L3 Cache (LLC)} & 12 MiB, 12 ways \\
        \textbf{Memory (Capacity / Speed)} & 128 GiB / 2666 MT/s \\
        \textbf{Storage} & 1 TB NVME SSD \\
        \textbf{Network} & 10 Gbps \\
        \bottomrule
    \end{tabular}
    \label{tab:platform}
\end{table}

\subsection{Workloads}\label{sec:workloads}

\begin{figure*}
    \centering
    \scalebox{.5}{
        \fontsize{14}{14}\selectfont 
        \includesvg{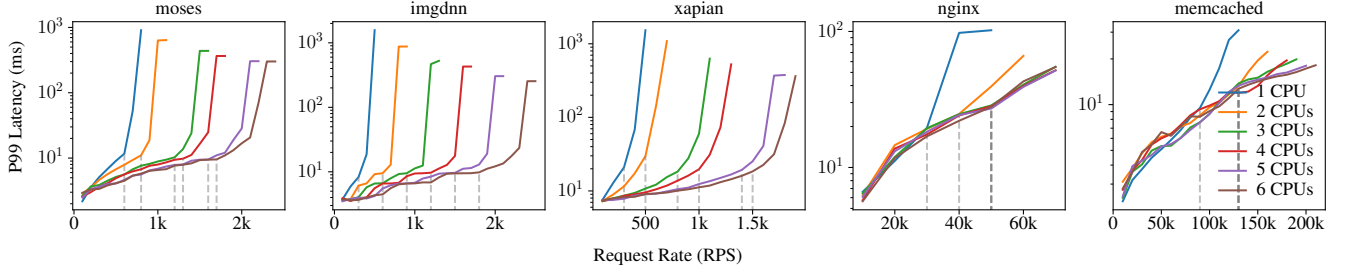}
    }
    \caption{Sizing experiments for different workloads. We see clear differences in performance between 
    core allocations for Moses, Img-dnn, and Xapian. In the case of Nginx and Memcached, they do not benefit 
    from additional CPU allocation above 2 cores, indicating that they are bottlenecked by other resources. It is worth noting
    the effects of core pinning and simultaneous multithreading that are visible in the Moses and Xapian cases, where performance increases are
    larger when going from an even to an odd number of cores as another physical core is engaged.}
    \label{fig:workloads}
\end{figure*}

We used a diverse set of workloads in our emulation environment, covering 
various application domains and resource usage patterns. 
\tablename~\ref{tab:workloads} summarizes the workloads. Moses, Img-dnn, and Xapian 
were adapted from the Tailbench benchmark suite~\cite{kastureTailbenchBenchmarkSuite2016} and
provide representative data-driven applications with latency requirements.
Nginx\footnote{\url{https://nginx.org}} and Memcached\footnote{\url{https://memcached.org}} 
are widely used latency critical applications with a lighter, more cache-sensitive resource profile.
Together, these workloads represent a broad spectrum of modern cloud applications.

\begin{table}[tbp]
    \centering
    \caption{Description of workloads.}
    \footnotesize
    \begin{tabular}{ll}
        \toprule
        \textbf{Workload} & \textbf{Description} \\ \midrule
                          Moses & Statistical machine translation \\
                          Img-dnn & Image recognition \\
                          Xapian & Online search \\
                          Nginx & Web server \\
                          Memcached & Key-value store \\
        \bottomrule
    \end{tabular}
    \label{tab:workloads}
\end{table}

Load was generated using a custom open-loop load generator that
records request latencies and success rates and exports these metrics to Prometheus.
For Moses, Img-dnn, and Xapian, we used the same load configurations, 
i.e. problem space and request patterns, as in Tailbench~\cite{kastureTailbenchBenchmarkSuite2016}.
The Nginx web server hosted $10^6$ static 1KB HTML pages that were requested using HTTP GET requests, 
while the Memcached key-value store was loaded with $10^5$ key-value pairs with 
a value size of 200B that were similarly requested using get commands.
The request distribution to both the Nginx and Memcached services were generated using a Zipf distribution.

In order to properly dimension the workloads and their resource demands, we
performed a series of sizing experiments. Each workload was executed on the node
under test in isolation while varying its CPU allocation and load intensity.
This allowed us to identify ``knees'' in the performance curves similarly to previous 
studies, where additional load leads to significant performance degradation~\cite{patelCLITEEfficientQoSAware2020,chenPARTIESQoSAwareResource2019}. 
These results were then used to
determine load levels and performance targets (SLOs) for each workload during 
the main experiments. Specifically, we selected the load as 80\% of the knee load, 
and set the SLO as 150\% of the measured $99^{\textrm{th}}$ percentile tail latency at 
the knee load. See \figurename~\ref{fig:workloads} for results.

\subsection{Experiment Design} \label{sec:experiment-design}

Even though workload scheduling and re-scheduling is a cluster-level problem, 
the core of our approaches focus on node-by-node mechanisms and resource awareness. 
Thus, to better understand and evaluate the effectiveness of our approach, we first observed
a single node in isolation. 
In a real world scenario, the workloads of a cluster would be
distributed across the cluster's available worker nodes. 
However, any workload not scheduled on the node under evaluation 
is assumed to be placed somewhere else and thus ignored.

There are two main components we evaluate, each at a different time in the
Pod lifecycle. First, we evaluate the filtering component at scheduling time. 
This component is responsible for admission control, 
i.e. evaluating the feasibility of scheduling a Pod on the node
given the current resource usage and availability. Second, we evaluate the
descheduling component, responsible for identifying needed rescheduling actions 
when the node becomes overloaded, or resources become stranded. Identified
Pods are then evicted, triggering the cluster's scheduler to re-schedule to other
nodes as needed.

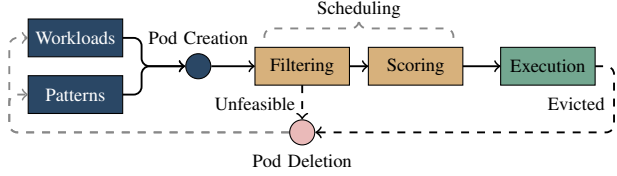
\begin{figure}[tbp]
    \centering
    \begin{tikzpicture}
	\begin{pgfonlayer}{nodelayer}
		\node [style=Small Circle] (0) at (-1.5, 0) {};
		\node [style=none] (1) at (-3, 0) {};
		\node [style=none] (2) at (-3, 0.75) {};
		\node [style=none] (3) at (-3, -0.75) {};
		\node [style=none] (4) at (-3.5, -0.75) {};
		\node [style=none] (5) at (-3.5, 0.75) {};
		\node [style=none] (8) at (-6, 0.25) {};
		\node [style=none] (9) at (-6, 1.25) {};
		\node [style=none] (10) at (-3.5, 1.25) {};
		\node [style=none] (11) at (-3.5, 0.25) {};
		\node [style=none] (14) at (-6, -1.25) {};
		\node [style=none] (15) at (-6, -0.25) {};
		\node [style=none] (16) at (-3.5, -0.25) {};
		\node [style=none] (17) at (-3.5, -1.25) {};
		\node [style=none] (19) at (2.5, 0) {};
		\node [style=none] (21) at (0, -0.5) {};
		\node [style=none] (22) at (0, 0.5) {};
		\node [style=none] (23) at (2.5, 0.5) {};
		\node [style=none] (24) at (2.5, -0.5) {};
		\node [style=none] (25) at (0, 0) {};
		\node [style=none] (26) at (5.5, 0) {};
		\node [style=none] (28) at (3, -0.5) {};
		\node [style=none] (29) at (3, 0.5) {};
		\node [style=none] (30) at (5.5, 0.5) {};
		\node [style=none] (31) at (5.5, -0.5) {};
		\node [style=none] (32) at (3, 0) {};
		\node [style=none] (35) at (9, 0) {};
		\node [style=none] (37) at (6.5, -0.5) {};
		\node [style=none] (38) at (6.5, 0.5) {};
		\node [style=none] (39) at (9, 0.5) {};
		\node [style=none] (40) at (9, -0.5) {};
		\node [style=none] (41) at (6.5, 0) {};
		\node [style=none] (42) at (0.25, 1) {};
		\node [style=none] (43) at (2.75, 1) {};
		\node [style=none] (44) at (2.75, 1.25) {};
		\node [style=none] (45) at (5.25, 1) {};
		\node [style=none] (46) at (0.25, 0.75) {};
		\node [style=none] (47) at (5.25, 0.75) {};
		\node [style=none] (51) at (9.5, 0) {};
		\node [style=none] (52) at (9.5, -1.75) {};
		\node [style=none] (53) at (-6, 0.75) {};
		\node [style=none] (54) at (-6.5, 0.75) {};
		\node [style=none] (55) at (-6.5, -0.75) {};
		\node [style=none] (56) at (-6, -0.75) {};
		\node [style=none] (57) at (-6.5, -1.75) {};
		\node [style=none] (59) at (1.25, -0.5) {};
		\node [style=Small Circle, fill={rgb,255: red,234; green,186; blue,185}] (49) at (1.25, -1.75) {};
		\node [style=Text, font={\scriptsize}] (48) at (2.75, 1.5) {Scheduling};
		\node [style=White Text, font={\scriptsize}] (6) at (-4.75, 0.75) {Workloads};
		\node [style=Text, font={\scriptsize}] (18) at (-1.5, 0.75) {Pod Creation};
		\node [style=Text, font={\scriptsize}] (20) at (1.25, 0) {Filtering};
		\node [style=Text, font={\scriptsize}] (27) at (4.25, 0) {Scoring};
		\node [style=Text, font={\scriptsize}] (36) at (7.75, 0) {Execution};
		\node [style=White Text, font={\scriptsize}] (7) at (-4.75, -0.75) {Patterns};
		\node [style=Text, font={\scriptsize}] (58) at (0, -1) {Unfeasible};
		\node [style=Text, font={\scriptsize}] (60) at (8.5, -1) {Evicted};
		\node [style=Text, font={\scriptsize}] (50) at (1.25, -2.5) {Pod Deletion};
	\end{pgfonlayer}
	\begin{pgfonlayer}{edgelayer}
		\draw [style=Arrow, rounded corners=3] (4.center)
			 to (3.center)
			 to (1.center)
			 to (0);
		\draw [style=Arrow, rounded corners=3] (5.center)
			 to (2.center)
			 to (1.center)
			 to (0);
		\draw [style=Freebox] (8.center)
			 to (9.center)
			 to (10.center)
			 to (11.center)
			 to cycle;
		\draw [style=Freebox] (14.center)
			 to (15.center)
			 to (16.center)
			 to (17.center)
			 to cycle;
		\draw [style=Freebox Yellow] (21.center)
			 to (22.center)
			 to (23.center)
			 to (24.center)
			 to cycle;
		\draw [style=Arrow] (0) to (25.center);
		\draw [style=Freebox Yellow] (28.center)
			 to (29.center)
			 to (30.center)
			 to (31.center)
			 to cycle;
		\draw [style=Arrow] (19.center) to (32.center);
		\draw [style=Freebox Green] (39.center)
			 to (40.center)
			 to (37.center)
			 to (38.center)
			 to cycle;
		\draw [style=Dashed, rounded corners=3] (46.center)
			 to (42.center)
			 to (43.center)
			 to (44.center);
		\draw [style=Dashed, rounded corners=3] (47.center)
			 to (45.center)
			 to (43.center)
			 to (44.center);
		\draw [style=Dashed Arrow, rounded corners=3] (35.center)
			 to (51.center)
			 to (52.center)
			 to (49);
		\draw [style=Thin Dashed Arrow, rounded corners=3] (49)
			 to (57.center)
			 to (54.center)
			 to (53.center);
		\draw [style=Thin Dashed Arrow, rounded corners=3] (49)
			 to (57.center)
			 to (55.center)
			 to (56.center);
		\draw [style=Arrow] (26.center) to (41.center);
		\draw [style=Dashed Arrow, rounded corners=3] (59.center) to (49);
	\end{pgfonlayer}
\end{tikzpicture}
    \caption{Illustration of experiment cycle. New Pods are created periodically using a random
        combination of workload type, size, and load pattern. It then passes through filtering and scoring
        before eventually starting execution should schedule be successful. It remains running on the node unless
    it is evicted by the descheduler, at which point it is terminated and deleted.}
    \label{fig:cycle}
\end{figure}

To emulate new Pods being created in the cluster, due to workload scaling
or new workload arrivals, we periodically create a new Pod at random 
from the set of defined workloads. This Pod then goes through the scheduling
process, where our filtering component is evaluated. Should the node not be
considered feasible for the Pod, it is simply discarded. Otherwise, it is
scheduled on the node, and starts executing. Over time, as workloads are
scheduled on the node, resource contention may arise, triggering the
descheduling component and evictions as needed. This cycle is illustrated in
\figurename~\ref{fig:cycle}.

Each workload is created with a random load pattern to emulate different load
variations in real-world workloads. 
The patterns used
during our experiments are shown in \figurename~\ref{fig:modifiers}. 
Repeating this experiment multiple times with different random seeds allows us to
gather a sufficient amount of data to evaluate our approach. 
We initially perform a baseline experiment where no resource-aware scheduling is performed.
I.e. Pods are scheduled as long as sufficient resources are available,
regardless of the actual resource usage patterns.

\begin{figure}[tbp]
    \centering
    \scalebox{.5}{
        \fontsize{14}{14}\selectfont 
        \includesvg{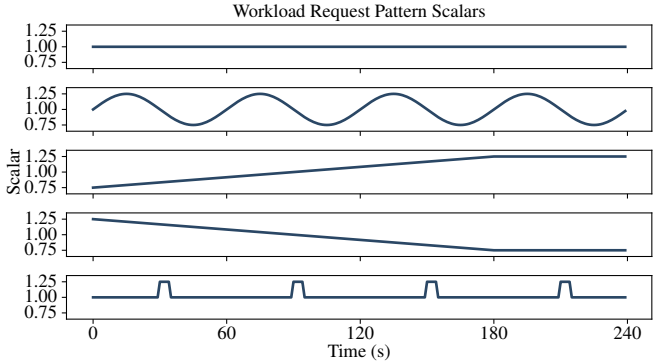}
    }
    \caption{Workload request patterns used in experiments. The scalar of the pattern is 
        multiplied with the base load of the workload to create varying load levels over time.
        Each pattern is repeated every 240 seconds.
    }\label{fig:modifiers}
\end{figure}

\subsection{Simulation Environment} \label{sec:simulation-setup}

In addition to the single-node experiments, we also created a virtual cluster simulation
environment to evaluate our scheduling strategy in a multi-node context, as inspired by evaluation methods
of similar studies~\cite{zambiancoDisruptionAwareMicroserviceReOrchestration2025, marcheseEnhancingKubernetesPlatform2025}. 
This simulation is implemented on top of \texttt{kind},\footnote{\url{https://github.com/kubernetes-sigs/kind}}
a Kubernetes-in-Docker solution that allows us to create a virtual Kubernetes cluster
using containerized nodes. The benefits of this approach are twofold. First, it allows us to
easily create and manage multi-node cluster environments of varying sizes.
Second, it allows us to run real Kubernetes components, including our custom scheduler
and descheduler, in our simulated environment.

To achieve this, we create \texttt{kind} clusters with a single control-plane node and 
varying numbers of worker nodes.
We then deploy our custom scheduler and descheduler components to the control-plane node,
together with Prometheus which is used for monitoring and data collection.
We then simulate the experiment through the use of mock workloads. Each workload
is represented as a lightweight Pod that periodically reports fictive metrics on resource usage
and other values of interest to Prometheus. The reported values are generated based
on metrics collected from the single-node experiments, ensuring that the simulated workloads
behave similarly to their real-world counterparts. Gaussian noise with a standard deviation
of 5\% is added to the reported values to emulate real-world variability. \figurename~\ref{fig:simulation}
shows an overview of both the emulation and simulation environments and their differences.

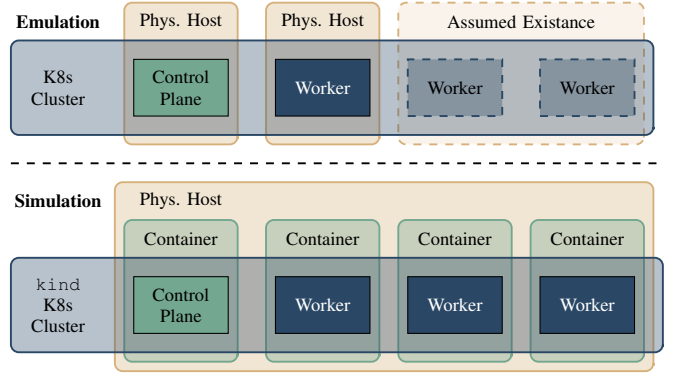
\begin{figure}[tbp]
    \centering
    \begin{tikzpicture}
	\begin{pgfonlayer}{nodelayer}
		\node [style=none] (0) at (-5.75, 1.5) {};
		\node [style=none] (1) at (-5.75, 0) {};
		\node [style=none] (2) at (-3.25, 0) {};
		\node [style=none] (3) at (-3.25, 1.5) {};
		\node [style=none] (4) at (-2, 1.5) {};
		\node [style=none] (5) at (-2, 0) {};
		\node [style=none] (6) at (0.5, 0) {};
		\node [style=none] (7) at (0.5, 1.5) {};
		\node [style=none] (8) at (1.5, 1.5) {};
		\node [style=none] (9) at (1.5, 0) {};
		\node [style=none] (10) at (4, 0) {};
		\node [style=none] (11) at (4, 1.5) {};
		\node [style=none] (12) at (5, 1.5) {};
		\node [style=none] (13) at (5, 0) {};
		\node [style=none] (14) at (7.5, 0) {};
		\node [style=none] (15) at (7.5, 1.5) {};
		\node [style=none] (16) at (-9, 2) {};
		\node [style=none] (17) at (-9, -0.5) {};
		\node [style=none] (18) at (8, -0.5) {};
		\node [style=none] (19) at (8, 2) {};
		\node [style=none] (25) at (-6, 3) {};
		\node [style=none] (26) at (-6, -0.75) {};
		\node [style=none] (27) at (-3, -0.75) {};
		\node [style=none] (28) at (-3, 3) {};
		\node [style=none] (30) at (-2.25, 3) {};
		\node [style=none] (31) at (-2.25, -0.75) {};
		\node [style=none] (32) at (0.75, -0.75) {};
		\node [style=none] (33) at (0.75, 3) {};
		\node [style=none] (35) at (1.25, 3) {};
		\node [style=none] (36) at (1.25, -0.75) {};
		\node [style=none] (37) at (7.75, -0.75) {};
		\node [style=none] (38) at (7.75, 3) {};
		\node [style=none] (45) at (-5.75, -4.25) {};
		\node [style=none] (46) at (-5.75, -5.75) {};
		\node [style=none] (47) at (-3.25, -5.75) {};
		\node [style=none] (48) at (-3.25, -4.25) {};
		\node [style=none] (49) at (-2, -4.25) {};
		\node [style=none] (50) at (-2, -5.75) {};
		\node [style=none] (51) at (0.5, -5.75) {};
		\node [style=none] (52) at (0.5, -4.25) {};
		\node [style=none] (53) at (1.5, -4.25) {};
		\node [style=none] (54) at (1.5, -5.75) {};
		\node [style=none] (55) at (4, -5.75) {};
		\node [style=none] (56) at (4, -4.25) {};
		\node [style=none] (57) at (5, -4.25) {};
		\node [style=none] (58) at (5, -5.75) {};
		\node [style=none] (59) at (7.5, -5.75) {};
		\node [style=none] (60) at (7.5, -4.25) {};
		\node [style=none] (61) at (-9, -3.75) {};
		\node [style=none] (62) at (-9, -6.25) {};
		\node [style=none] (63) at (8.25, -6.25) {};
		\node [style=none] (64) at (8.25, -3.75) {};
		\node [style=none] (70) at (-6, -2.75) {};
		\node [style=none] (71) at (-6, -6.5) {};
		\node [style=none] (72) at (-3, -6.5) {};
		\node [style=none] (73) at (-3, -2.75) {};
		\node [style=none] (75) at (-2.25, -2.75) {};
		\node [style=none] (76) at (-2.25, -6.5) {};
		\node [style=none] (77) at (0.75, -6.5) {};
		\node [style=none] (78) at (0.75, -2.75) {};
		\node [style=none] (80) at (1.25, -2.75) {};
		\node [style=none] (81) at (1.25, -6.5) {};
		\node [style=none] (82) at (4.25, -6.5) {};
		\node [style=none] (83) at (4.25, -2.75) {};
		\node [style=none] (85) at (4.75, -2.75) {};
		\node [style=none] (86) at (4.75, -6.5) {};
		\node [style=none] (87) at (7.75, -6.5) {};
		\node [style=none] (88) at (7.75, -2.75) {};
		\node [style=none] (90) at (8.25, -1.25) {};
		\node [style=none] (91) at (-9, -1.25) {};
		\node [style=none] (92) at (-6.25, -1.75) {};
		\node [style=none] (93) at (-6.25, -6.75) {};
		\node [style=none] (94) at (8, -6.75) {};
		\node [style=none] (95) at (8, -1.75) {};
		\node [style=Text, font={\scriptsize}] (97) at (-7.75, 2.5) {\textbf{Emulation}};
		\node [style=Text, font={\scriptsize}] (29) at (-4.5, 2.5) {Phys. Host};
		\node [style=Text, font={\scriptsize}] (34) at (-0.75, 2.5) {Phys. Host};
		\node [style=Text, font={\scriptsize}] (39) at (4.5, 2.5) {Assumed Existance};
		\node [style=Text, font={\scriptsize}] (20) at (-7.75, 0.75) {K8s\\Cluster};
		\node [style=Text, font={\scriptsize}] (21) at (-4.5, 0.75) {Control\\Plane};
		\node [style=White Text, font={\scriptsize}] (22) at (-0.75, 0.75) {Worker};
		\node [style=Text, font={\scriptsize}] (23) at (2.75, 0.75) {Worker};
		\node [style=Text, font={\scriptsize}] (24) at (6.25, 0.75) {Worker};

		\node [style=Text, font={\scriptsize}] (98) at (-7.75, -2.25) {\textbf{Simulation}};
		\node [style=Text, font={\scriptsize}] (96) at (-4.5, -2.25) {Phys. Host};
		\node [style=Text, font={\scriptsize}] (74) at (-4.5, -3.25) {Container};
		\node [style=Text, font={\scriptsize}] (79) at (-0.75, -3.25) {Container};
		\node [style=Text, font={\scriptsize}] (84) at (2.75, -3.25) {Container};
		\node [style=Text, font={\scriptsize}] (89) at (6.25, -3.25) {Container};
        \node [style=Text, font={\scriptsize}] (65) at (-7.75, -5) {\texttt{kind}\\K8s\\Cluster};
		\node [style=Text, font={\scriptsize}] (66) at (-4.5, -5) {Control\\Plane};
		\node [style=White Text, font={\scriptsize}] (67) at (-0.75, -5) {Worker};
		\node [style=White Text, font={\scriptsize}] (68) at (2.75, -5) {Worker};
		\node [style=White Text, font={\scriptsize}] (69) at (6.25, -5) {Worker};
	\end{pgfonlayer}
	\begin{pgfonlayer}{edgelayer}
		\draw [style=Yellow EC F, rounded corners=3] (27.center)
			 to (28.center)
			 to (25.center)
			 to (26.center)
			 to cycle;
		\draw [style=Yellow EC F, rounded corners=3] (32.center)
			 to (33.center)
			 to (30.center)
			 to (31.center)
			 to cycle;
		\draw [style=Yellow EC F, rounded corners=3, dashed, fill opacity=0.15] (37.center)
			 to (38.center)
			 to (35.center)
			 to (36.center)
			 to cycle;
		\draw [style=Blue EC F, rounded corners=3] (18.center)
			 to (19.center)
			 to (16.center)
			 to (17.center)
			 to cycle;
		\draw [style=Yellow EC F, rounded corners=3] (94.center)
			 to (95.center)
			 to (92.center)
			 to (93.center)
			 to cycle;
		\draw [style=Freebox Green] (1.center)
			 to (2.center)
			 to (3.center)
			 to (0.center)
			 to cycle;
		\draw [style=Freebox] (5.center)
			 to (6.center)
			 to (7.center)
			 to (4.center)
			 to cycle;
		\draw [style=Blue EC F, dashed] (9.center)
			 to (10.center)
			 to (11.center)
			 to (8.center)
			 to cycle;
		\draw [style=Blue EC F, dashed] (13.center)
			 to (14.center)
			 to (15.center)
			 to (12.center)
			 to cycle;
		\draw [style=Green EC F, rounded corners=3] (72.center)
			 to (73.center)
			 to (70.center)
			 to (71.center)
			 to cycle;
		\draw [style=Green EC F, rounded corners=3] (77.center)
			 to (78.center)
			 to (75.center)
			 to (76.center)
			 to cycle;
		\draw [style=Green EC F, rounded corners=3] (82.center)
			 to (83.center)
			 to (80.center)
			 to (81.center)
			 to cycle;
		\draw [style=Green EC F, rounded corners=3] (87.center)
			 to (88.center)
			 to (85.center)
			 to (86.center)
			 to cycle;
		\draw [style=Blue EC F, rounded corners=3] (63.center)
			 to (64.center)
			 to (61.center)
			 to (62.center)
			 to cycle;
		\draw [style=Freebox Green] (46.center)
			 to (47.center)
			 to (48.center)
			 to (45.center)
			 to cycle;
		\draw [style=Freebox] (50.center)
			 to (51.center)
			 to (52.center)
			 to (49.center)
			 to cycle;
		\draw [style=Freebox] (54.center)
			 to (55.center)
			 to (56.center)
			 to (53.center)
			 to cycle;
		\draw [style=Freebox] (58.center)
			 to (59.center)
			 to (60.center)
			 to (57.center)
			 to cycle;
		\draw [style=Thick Dashed, in=180, out=0] (91.center) to (90.center);
	\end{pgfonlayer}
\end{tikzpicture}
    \caption{Overview of the real-world emulation and cluster simulation environments.
        The simulation extends the emulation by creating multiple worker nodes
        using \texttt{kind}, and using mock workloads that report fictive resource usage
        based on data collected from the emulation environment.}
    \label{fig:simulation}
\end{figure}

\section{Evaluation}

In order to ensure a comprehensive evaluation of our proposed solutions,
we ask a series of questions that we aim to answer through our experiments:
\begin{enumerate}
    \item \label{q1} Does our SLO-based rescheduling approach improve SLO compliance
        without major reductions in resource utilization while treating SLOs as hard limits?
    \item \label{q2} Can soft SLO limits effectively reduce the rescheduling actions
        required compared to a hard-limit approach?
    \item \label{q3} How do different lookback window sizes and eviction thresholds for soft SLO limits
        affect the performance of our soft-limit approach?
    \item \label{q4} Does buoyancy-aware scheduling reduce resource contention
        and improve SLO compliance compared to a baseline Kubernetes scheduler?
    \item \label{q5} How does buoyancy-aware scheduling perform in larger
        cluster environments compared to a single-node setup?
\end{enumerate}

We set out to answer questions~\ref{q1}, \ref{q2}, and~\ref{q3} in Section~\ref{sec:soft-limits-evaluation}, 
and questions~\ref{q4} and~\ref{q5} in Section~\ref{sec:buoyancy-evaluation}.

\subsection{Soft Limits and Margins}\label{sec:soft-limits-evaluation}

To evaluate the effectiveness of soft SLO limits with margins, we performed
the emulation experiment under different
lookback window sizes $w$. For simplicity and ease of comparison, we treat the 
lookback window $w$ as a window in wall clock time. This is later mapped to
discrete time steps based on the sampling interval used for monitoring 
workload performance and computing SLO slack. Workloads that have 
an average-over-time slack $\overline{s} < 0$ were evicted, reducing node interference 
and making space for other workloads on the node.
We compare different lookback sizes in \figurename~\ref{fig:slomargin}.

\begin{figure}[tbp]
    \centering
    \scalebox{.5}{
        \fontsize{14}{14}\selectfont 
        \includesvg{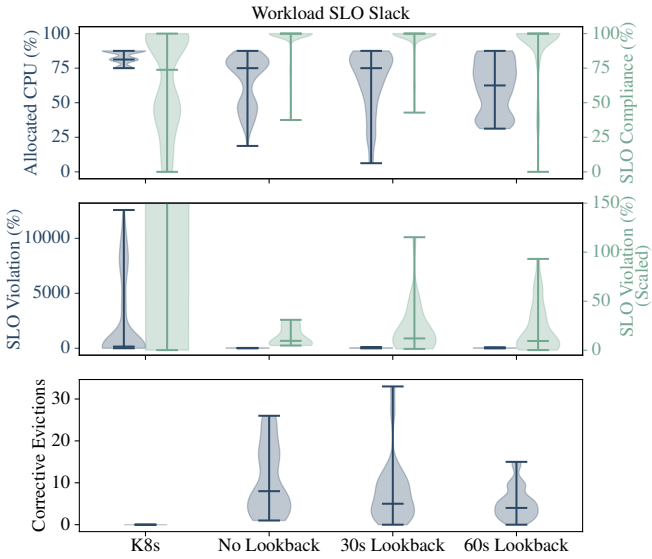}
    }
    \caption{Comparison of scheduling using different lookback window sizes for 
        SLO soft limits. This figure shows the aggregated results from 25 runs, 
        where each data point is a momentary snapshot of the system state during the experiment. 
        While the standard Kubernetes configuration tends to schedule a slightly
        larger fraction of workloads successfully, less than 75\% of workloads
        comply with their SLOs. Our approach using soft SLO limits with margins
        improves SLO compliance significantly, and in the events that workloads
        do not comply, the degree of violation is significantly reduced (by up to 
        two orders of magnitude) compared to the standard Kubernetes scheduler.
    }
    \label{fig:slomargin}
\end{figure}

Observing the results, we see that Kubernetes will generally schedule a slightly
larger fraction of workloads successfully compared to our approach, see \tablename~\ref{tab:slomarginresults}. 
While doing so, however, only 65\% of workloads comply with their SLOs on average, and when
they do not, the degree of violation is on average two orders of magnitude larger. 
In contrast, using our SLO-driven approach without lookback, a much greater 
fraction of workloads complies with their SLOs, and when they do not,
the degree of violation is greatly reduced.

\begin{table}[tbp]
    \centering
    \caption{Averages from SLO soft limit experiment comparing different lookback window sizes.
        Shows values from the same experiment as in \figurename~\ref{fig:slomargin}.}
    \footnotesize
    \begin{tabular}{l|rrrr}
        \toprule
        \textbf{Average} & \textbf{K8s}& \textbf{0s} & \textbf{30s}& \textbf{60s} \\ \midrule
        Allocated CPU (\%) & 83.00 & 65.88 & 65.60 & 61.33 \\ \midrule
        SLO Compliant (\%) & 65.18 & 98.18 & 97.48 & 92.25 \\\midrule
        Size of Violation (\%) & 2510 & 13.64 & 21.73 & 21.97 \\\midrule
        Evictions & 0.00 & 10.53 & 7.88 & 5.30  \\
        \bottomrule
    \end{tabular}
    \label{tab:slomarginresults}
\end{table}

Comparing lookback window sizes, we observe that a window size increase
allows the system to find steady states using fewer rescheduling actions.
Moving from no lookback to a 60-second window reduces the evictions
from $10.5$ to $5.3$ on average, a decrease of 49\%.
However, doing so also slightly increases the size of transient violations, and
reduces the total allocated workload count slightly. In this experiment, this
is due to the fact that as fewer rescheduling actions are taken, fewer 
workload combinations are explored, meaning the system may settle in a less
optimal state. Still, if we compare the allocated CPU values with the SLO
compliance ratios, we get that 54\% worth of system CPU is used 
by workloads in compliance on average by Kubernetes, while the worst
case using our approach (60s lookback) achieves 57\% CPU usage within 
SLO compliance. We achieve this while reducing the size
of an average SLO violation.

Furthermore, we performed additional experiments to explore the effects of using 
different eviction thresholds. Instead of evicting workloads when their average-over-time
slack $\overline{s} < 0$, we explored the effects of using other thresholds
(e.g., $\overline{s} < -0.1$, $\overline{s} < -0.2$) to further tune the sensitivity
of the rescheduling mechanism. The results are shown
in \figurename~\ref{fig:slotuning}.

\begin{figure}[tbp]
    \centering
    \scalebox{.5}{
        \fontsize{14}{14}\selectfont 
        \includesvg{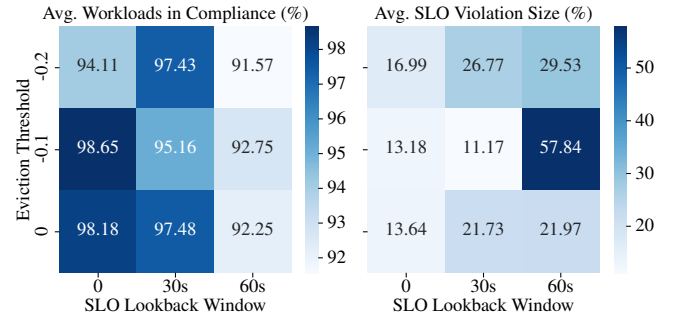}
    }
    \caption{Same experiment as in \figurename~\ref{fig:slomargin}, but exploring
        different lookback window sizes and eviction thresholds. The eviction
        threshold defines the value of average-over-time slack $\overline{s}$ below which
        workloads are evicted. Results are once again aggregated over 25 experiment runs.
    }
    \label{fig:slotuning}
\end{figure}

The experiments show that the ``$0s$'' lookback case performs best in terms of absolute
compliance. That is to be expected as it does not tolerate any transient violations. However,
we can also see that a lookback window of $30s$ achieves very similar compliance 
ratios while reducing the number of rescheduling actions required.
We may also note that adjusting the lookback size and the eviction threshold
have similar effects on the number of workloads that manages to comply with their SLOs. 
However, it is largely the size of the lookback window that determines the degree of SLO violation
when workloads do not comply, whereas the eviction threshold has less effect on this aspect.
This indicates that an increased lookback window indeed allows larger transient breaches
without triggering rescheduling actions and that significant and sustained breaches will trigger 
rescheduling actions eventually.

\subsubsection{Summary}

In evaluation of our soft-limit approach we have seen that using soft limits for SLOs 
is an effective strategy of maintaining SLO compliance while 
reducing the number of rescheduling actions required to do so compared to a 
hard limit approach. By tuning the lookback window size and eviction threshold,
system operators may further adjust the sensitivity of the rescheduling mechanism 
to transient SLO breaches, allowing them to find a suitable balance between
system stability and SLO compliance.

\subsection{Buoyancy Based Scheduling} \label{sec:buoyancy-evaluation}

We first evaluated our buoyancy-aware scheduling strategy in the single-node
experimental setup introduced in Section~\ref{sec:experiment-design}.
This emulation allows us to gain an understanding of how our approach
performs in isolation, without the added complexity of a full cluster environment.
While this setup does not capture the full dynamics of a multi-node cluster,
nor does it allow us to evaluate the effects of the scoring phase, it provides
valuable insights into the effectiveness of our filtering and descheduling mechanisms
in the context of individual nodes. Results are presented
in \figurename~\ref{fig:buoyancy}.

\begin{figure}[tbp]
    \centering
    \scalebox{.5}{
        \fontsize{14}{14}\selectfont 
        \includesvg{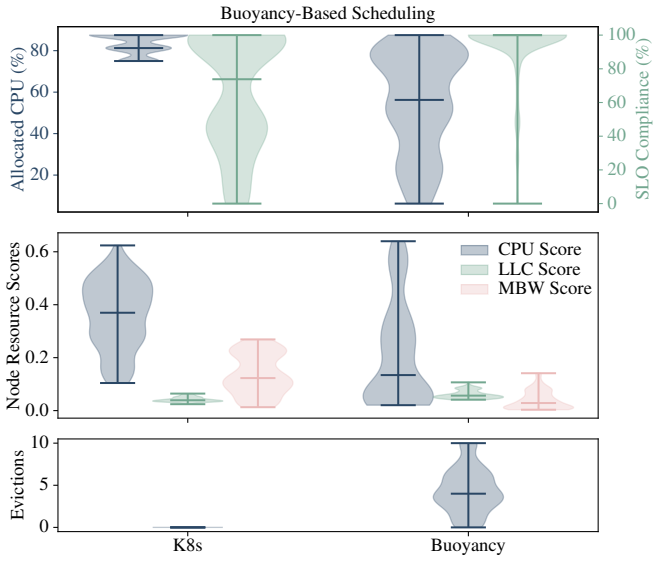}
    }
    \caption{Buoyancy-aware scheduling experiment results. 
        Results are aggregated from 25 experiment runs. We see that the buoyancy-aware
        scheduling strategy significantly reduces the number of SLO violations compared to 
        the Kubernetes baseline. This is achieved through a lower resource congestion
        overall, as indicated by the lower resource scores. This is achieved using just a few 
        corrective evictions.}
    \label{fig:buoyancy}
\end{figure}

From the results, we observe that buoyancy-aware scheduling reduces the total 
number of workloads admitted to the node compared to the Kubernetes baseline.
While doing so, however, it improves SLO compliance, with over 93\% of workloads
meeting their SLOs compared to just 65\% under the Kubernetes baseline.
Furthermore, we can observe the resource congestion, as represented by the node-level
resource scores, also decrease substantially in CPU and the memory bandwidth, while
slightly increasing in the LLC score. As LLC is the least contended resource throughout this experiment, 
it is expected that it will become more contended as CPU and memory 
bandwidth contention is reduced. On average, the resource scores showed a $0.08$
decrease, indicating a notable reduction in overall resource contention on the node.
This may be viewed as a decrease in total resource saturation by approximately 8\% of available resources.
As a final observation, we note that the buoyancy-aware scheduling
approach achieves these improvements using just a few corrective evictions.

While this experiment demonstrates buoyancy-aware scheduling in a single-node
context, it does not capture the benefits of the scoring phase of the scheduling process.

\subsubsection{Cluster Simulation}

To evaluate the full buoyancy-aware scheduling strategy, including scoring,
we additionally performed an evaluation using the simulated cluster environment. 
This simulation allows us to model a multi-node cluster and observe how
the benefits of buoyancy-aware scheduling manifest at the cluster level.
This experiment is repeated for cluster sizes of 1, 8, and 32 nodes to 
both validate the single-node results and observe how the approach scales.
The results of this experiment are presented in \figurename~\ref{fig:sim}.

\begin{figure}[tbp]
    \centering
    \scalebox{.5}{
        \fontsize{14}{14}\selectfont 
        \includesvg{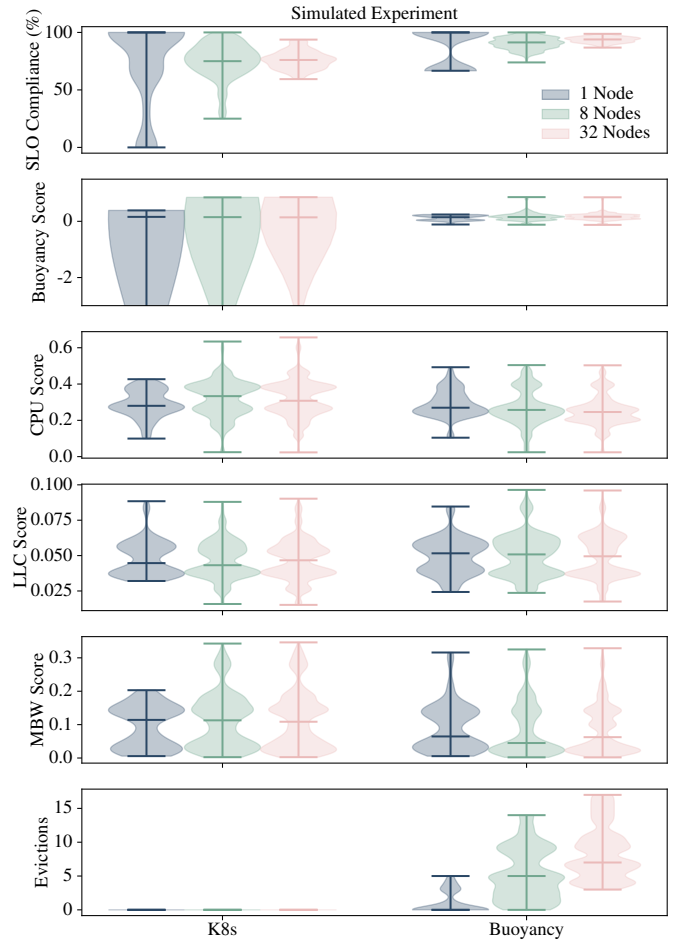}
    }
    \caption{Simulated experiment results for cluster sizes of 1, 8, and 32 nodes.
        As cluster size increases, the benefits of buoyancy-aware scheduling become 
        more pronounced as fewer workloads experience SLO violations, while 
        resource congestion through workload buoyancy remains comparable.}
    \label{fig:sim}
\end{figure}

In the simulated cluster environment, we can observe similar SLO compliance numbers in the single node case compared
to the emulated single-node experiment we saw in \figurename~\ref{fig:buoyancy}. Here, we may also observe
that as the cluster size increases, a larger proportion of workloads comply with their SLOs in both the baseline, and the buoyancy
case. This indicates that both scheduling strategies benefit from the increased flexibility
of a larger number of nodes available during scheduling. However, the buoyancy-aware approach, with its
resource contention insight, is able to take better advantage of this flexibility,
resulting in increased in SLO compliance in larger clusters.

Furthermore, we see similar resource contention behavior as the 
single-node experiment showed, across all cluster sizes.
This, in combination with greatly improved buoyancy scores indicates that the buoyancy-aware
scheduling strategy is effective at reducing resource contention across the cluster,
leading to improved workload performance.

\subsubsection{Summary}

In both the single-node emulation experiment and the multi-node cluster simulation,
we have demonstrated the effectiveness of our buoyancy-aware scheduling strategy.
Through the use of buoyancy metrics in filtering, scoring, and descheduling,
we have shown decreases in resource contention, resulting in improvements
in SLO compliance compared to the Kubernetes baseline.

\section{Limitations and Threats to Validity} \label{sec:limitations}

As it stands, our proposed approaches of using soft SLO limits and workload buoyancy 
provide a good basis for scheduling and rescheduling mechanisms on the operational level. Outside the
scope for this work, however, we acknowledge that there are several opportunities for further
exploration using these approaches in other contexts. For example, they may be used 
to improve and provide insights to workload scaling mechanisms, both horizontal and vertical.

Furthermore, workload buoyancy and resource scores provide insight into which resources
are most contended for a given node. In this paper, we have not considered the 
resource usage characteristics of a Pod during scheduling. Doing so would allow for more informed
decisions regarding workload placement based on the specific resource demands of that workload.
However, this would require some notion of a resource usage profile of the workload being scheduled,
which may not be available or feasible to obtain.

\subsection{Threats to Validity}

Throughout the evaluation of our proposed approaches, care was taken to ensure the
validity and reliability of our results. We acknowledge, however, that there are
potential threats, both external and internal, which we outline in this section.

\subsubsection{External Threats}

Our implementation and evaluation builds upon, and integrates with, several existing
software technologies. Cloud native technologies such as Kubernetes, Prometheus,
and Kind provide a solid foundation for our work and evaluation, but also introduce potential
risks to its long-term validity. As these technologies evolve, there is a risk that
our implementation may become incompatible or require significant modifications to
function correctly. Similarly, our evaluation relies upon extension points in
the Kubernetes scheduler and descheduler that may change in future versions,
potentially impacting the future applicability of our results.

\subsubsection{Internal Threats}

Our emulation-based evaluation approach introduces several potential internal threats
to validity. First, due to hardware availability limitations, 
our evaluation was conducted by observing a single worker node in isolation while making assumptions about 
other nodes in the cluster. This approach may not fully capture the complexities
of a multi-node cluster environment, potentially impacting the generalizability of our results.
However, we ensured the experiment was as realistic as possible within these constraints.

To reason about cluster-level behavior, we used simulation techniques to approximate
cluster dynamics based on observations from a single node. While we took care to
ensure the accuracy of these approximations, there is a risk that they may not fully
capture the nuances of real-world cluster behavior.

Finally, our evaluation was conducted in a controlled environment, which may not
fully reflect the variability and unpredictability of real-world deployments.

\section{Related Work}

\subsubsection{Application Aware Scheduling}

Many scheduling approaches have been proposed that leverage application-level
metrics and performance indicators to inform scheduling decisions. 
\citeauthor*{giannakopoulosPerformanceAwareSchedulingLoad2025}~\cite{giannakopoulosPerformanceAwareSchedulingLoad2025}
attempt to schedule and load-balance applications based on performance predictions. Similarly, 
\citeauthor*{carvalhoQoEAwareContainerScheduler2021}~\cite{carvalhoQoEAwareContainerScheduler2021}
improve scheduling decisions by predicting the quality of experience (QoE) of applications.
\citeauthor*{da_silva_qos-driven_2020}~\cite{da_silva_qos-driven_2020} propose a QoS-driven scheduler
that make use of workload priorities and application-level metrics to inform scheduling decisions.
They follow the service level agreement (SLA) penalty models of major cloud providers during scheduling.
Furthermore, Owl~\cite{tianOwlPerformanceawareScheduling2022} leverages application-level
performance profiling of hot function-as-a-service functions to improve 
workload placement and reduce congestion.

These works demonstrate the benefits of application-aware scheduling. However, they 
consider performance targets as strict requirements that should never be broken. 
We believe that by allowing more flexibility, schedulers can make better
decisions that lead to improved system utilization over time.

\subsubsection{Resource Aware Partitioning and Scheduling}

There is no shortage of resource management studies that demonstrate clear
improvements to workload performance by partitioning shared system resources~\cite{larssonHardwareLevelQoSEnforcement2025}. 
Partitioning of LLC and memory bandwidth have shown significant improvements through reduced
interference~\cite{chenPARTIESQoSAwareResource2019, chenOLPartOnlineLearning2023, tangThemisFairMemory2023}.
Even so, such insights have been left unexplored in the context of cluster scheduling.
Resource aware scheduling proposals have considered resource domains 
such as CPU, memory, network, and disk to great effect using different approaches. 
Zeus~\cite{zhangZeusImprovingResource2021a} and Rhythm~\cite{zhaoRhythmComponentdistinguishableWorkload2020}
leverage different workload priorities in combination with resource awareness to improve scheduling decisions.
Some use heuristics to guide scheduling decisions~\cite{marcheseEnhancingKubernetesPlatform2025,zambiancoDisruptionAwareMicroserviceReOrchestration2025}, while others
use learning-based approaches~\cite{jianDRSDeepReinforcement2024} to adapt to dynamic environments.
Nautilus~\cite{fuQoSAwareResourceEfficient2021} is one proposal that does consider 
resource contention during scheduling. However, similar to other works, the focus 
is on traditional resources such as CPU, memory, and I/O, not shared resources in the memory subsystem.
Zeus~\cite{zhangZeusImprovingResource2021a} is one exception to the lack of 
LLC integration as it partitions LLC to preemptively provide cache isolation. However, the approach
does not leverage insight into the actual LLC pressure of the application or node. 

There is a notable gap in the literature on incorporating memory 
subsystem congestion awareness into cluster scheduling, despite evidence that 
shared resource contention can significantly impact application performance.

\section{Conclusions}

In this paper, we have identified two key gaps in current workload scheduling
research. First, we highlighted an apparent lack of flexibility in existing
SLO-based scheduling approaches, leading to rigid and suboptimal resource
allocations to provide sufficient headroom to meet transient workload spikes.
We introduced a novel approach that tolerates short-term
SLO breaches in a ``fuse''-like manner, treating SLOs like soft limits. Using our approach, 
the size of the average SLO violation decreased by two orders of magnitude compared 
to a baseline Kubernetes configuration, while using 49\% fewer 
rescheduling actions compared to a hard SLO limit approach.

Second, we observed that existing approaches does not address or leverage 
insights from congestion in shared memory subsystem resources. 
To address this gap, we proposed a scheduling and rescheduling approach that
leverages insight into resource congestion through the use of workload
buoyancy metrics. By incorporating these metrics into scheduling decisions, we
reduced overall node resource congestion by 8\% and drastically reduced SLO violations.

\section*{Acknowledgements}

Funding for this project was provided in part
by the Knut and Alice Wallenberg Foundation under grant
KAW 2019.0352 and by the eSSENCE Programme under the
Swedish Government’s Strategic Research Initiative. 
Generative AI (including ChatGPT and Copilot) was used to 
improve language and readability in the preparation of this work. 
The authors remain solely responsible for its contents.

\printbibliography

\end{document}